\documentclass[twocolumn,showpacs]{revtex4}
\usepackage{graphicx}%
\usepackage{dcolumn}
\usepackage{amsmath}

\begin{document}
\title[Short Title]
{Generation of two-mode nonclassical motional states \\and a
Fredkin gate operation in a two-dimensional ion trap}
\author{Xu-Bo Zou, Jaewan Kim\footnote{On leave from Samsung Advanced
Institute of Technology}, and Hai-Woong Lee\footnote{E-mail
address: hwlee@laputa.kaist.ac.kr}} \affiliation{ Department of
Physics, Korea Advanced Institute of Science and Technology,
Taejon 305-701,Korea}
\date{\today}

\begin{abstract}
We present an efficient scheme to generate two-mode SU(2)
macroscopic quantum superposition (Schr\"odinger cat) states,
entangled number states and entangled coherent states for the
vibrational motion of an ion trapped in a two-dimensional
harmonic potential well. We also show that the same scheme can be
used to realize a Fredkin gate operation.
\end{abstract}
\maketitle

The generation of nonclassical states has been widely studied in
the past both in theory and in experiment. The first significant
advances were made in quantum optics with a successful
experimental demonstration of antibunched light\cite{Paul} and
squeezed light\cite{Loudon}. Various optical schemes of generating
macroscopic quantum superposition (Schr\"{o}dinger cat) states
have also been studied\cite{YSB}, which led to an experimental
realization in a quantized cavity field\cite{Brune}. Recently,
possible ways of generating various two-mode entangled field
states have been proposed. For example, it has been
shown\cite{Sanders} that entangled coherent state, which is
obtained when two-mode coherent states are
superposed\cite{Sanders,Chai} and may thus be considered as a
two(multi)-mode generalization of single-mode Schr\"{o}dinger cat
states\cite{Manko}, can be produced using the nonlinear
Mach-Zehnder interferometer. A method to generate another type of
two-mode Schr\"{o}dinger cat states known as SU(2)
Schr\"{o}dinger cat state, which results when two different SU(2)
coherent states\cite{Buzek,Sanders2,GGJMO} are superposed, has
also been proposed\cite{Sanders2,GGJMO}. It has also been shown
that two-mode entangled number states can be generated using
nonlinear optical interactions\cite{Gerry}, which may then be
used to obtain the maximum sensitivity in phase measurements set
by the Heisenberg limit\cite{Bollinger} . In general, however,
experimental realization of nonclassical field states is
difficult, because quantum coherence characterizing these states
is easily destroyed by interaction with the environment.

Recent advances in ion cooling and trapping have opened new
prospects in nonclassical state generation. An ion confined in an
electromagnetic trap can be approximately described as a particle
in a harmonic potential with its center of mass (c.m.) exhibiting
quantum-mechanical simple harmonic motion. By appropriately
driving the ion with laser fields, its internal and external
degrees of freedom can be coupled to the extent that its
center-of-mass motion can be manipulated with relative ease. One
advantage of the trapped-ion system is that the decoherence effect
is relatively weak due to the extremely weak coupling between the
vibrational modes and the environment. It was realized that this
advantage of the trapped ion system makes it a promising candidate
for constructing quantum logic gates for quantum
computation\cite{CZ} as well as for producing nonclassical states
of the center-of-mass vibrational motion. In fact, the
controlled-NOT gate has been experimentally realized with trapped
ions in a linear chain\cite{MonroePRL}. At the same time
single-mode nonclassical states, such as Fock states, squeezed
states and Schr\"{o}dinger cat states, of the ion's vibrational
motion have been theoretically investigated\cite{CCVPZ,Gerry2} and
experimentally realized\cite{MM} with the trapped ion system. Most
recently, various schemes of producing two-mode nonclassical
states of the vibrational motion have been proposed using ions
confined in a two-dimensional trap\cite{Gerry2,GN,GGM}.

In this paper, we propose an efficient scheme to generate two-mode
SU(2) Schr\"{o}dinger cat states, entangled number states and
entangled coherent states of the vibrational motion of an ion in a
two-dimensional harmonic potential. We also show that the same
scheme can be used to construct a Fredkin gate
(controlled-exchange gate)\cite{Fredkin}. The optical Fredkin
gate can be built using two beam splitters and a Kerr
medium\cite{Milburn}. It has been shown that it needs six
two-qubit operations to construct a Fredkin gate\cite{HF}. The
optical Fredkin gate has given a simple implementation of the
quantum computer to solve Deutsch's problem\cite{CY}. Note that,
using dual-rail representation\cite{CY2,CY3}, the optical Fredkin
gate corresponds to a controlled-NOT gate; in this sense, the
optical Fredkin gate can be used to synthesize any quantum
computing devices\cite{CY,CY3}.

Our scheme relies on the interactions
\begin{equation}
H_1=\hbar\Omega_1(a^{\dagger}b+ab^{\dagger})\sigma_x
\end{equation}
and
\begin{equation}
H_2=\hbar\Omega_2(a^{\dagger}b+ab^{\dagger})\;,
\end{equation}
where $(a,a^{\dagger})$ and $(b,b^{\dagger})$ are the boson
annihilation and creation operators for the two c.m. modes in the
$x$ and $y$ directions, respectively, $\sigma_x$ is the Pauli
spin matrix operating on the internal two levels of the ion, and
$\Omega_1$ and $\Omega_2$ represent the strength of the respective
interactions. The interaction (1) can be realized in a trapped ion
system in a two-dimensional potential in the Lamb-Dicke limit by
applying two symmetrically positioned pairs of Raman lasers, with
the first pair tuned to the first red sidebands of the ion
vibration in the $x$ and $y$ directions, respectively, and the
second pair tuned to the first blue sidebands\cite{Steinbach}. The
interaction (2) gives parametric coupling of the two vibrational
modes and has already been employed by various authors for the
purpose of manipulating the ion system\cite{Bollinger,GN,Heinzen}.

We first consider the situation in which the ion is prepared in
the ground state $|\downarrow\rangle$ and the two c.m. modes in
the Fock states $|0\rangle_a$ and $|n\rangle_b$, respectively,
\begin{equation}
|\Psi(t=0)\rangle=|\downarrow\rangle|0\rangle_a|n\rangle_b \;.
\end{equation}
If laser pulses of interaction (1) is applied, the
system evolves to
\begin{eqnarray}
|\Psi(t)\rangle=&\dfrac{1}{2}\exp[-i\Omega_1t(a^{\dagger}b+ab^{\dagger})]
(|\downarrow\rangle+|\uparrow\rangle)|0\rangle_a|n\rangle_b
\nonumber\\
&+\dfrac{1}{2}\exp[i\Omega_1t(a^{\dagger}b+ab^{\dagger})]
(|\downarrow\rangle-|\uparrow\rangle)|0\rangle_a|n\rangle_b\;. \nonumber\\
\end{eqnarray}
Upon detection of the ion state, the state vector is projected
into
\begin{eqnarray}
|\Psi_{\pm}\rangle=&\exp[-i\Omega_1t(a^{\dagger}b+ab^{\dagger})]
|0\rangle_a|n\rangle_b \nonumber\\
&\pm
\exp[i\Omega_1t(a^{\dagger}b+ab^{\dagger})]|0\rangle_a|n\rangle_b\;,
\end{eqnarray}
where $|\Psi_+\rangle$ is obtained if the state
$|\downarrow\rangle$ of the ion is detected, and $|\Psi_-\rangle$
if the upper state $|\uparrow\rangle$ is detected. The state (5)
is the SU(2) Schr\"{o}dinger cat state of the form
$(|\zeta,j\rangle \pm |-\zeta,j\rangle)$\cite{GGJMO}, where
$|\zeta,j\rangle= \exp[\beta J_+ -\beta^* J_- ]|j,-j\rangle$ is
the SU(2) coherent state\cite{Buzek}, if the identification is
made as $\beta=-i \Omega_1 t$, $j=n/2$, and $\zeta=-i
\tan(\Omega_1 t/2)$.

If the interactions (1) and (2) are applied in succession to the
initial state (3) and the durations $t_1$ and $t_2$ of the
respective interactions are chosen to be
$\Omega_1t_1=\Omega_2t_2=\pi/4$, the system evolves to
\begin{eqnarray}
|\Psi(t=t_1 +t_2)\rangle=&\dfrac{1}{2}
i^n(|\downarrow\rangle+|\uparrow\rangle)
|n\rangle_a|0\rangle_b \nonumber\\
&+\dfrac{1}{2}
(|\downarrow\rangle-|\uparrow\rangle)|0\rangle_a|n\rangle_b\;.
\end{eqnarray}
Upon detection of the ion state, the state (6) collapses to
\begin{equation}
|\Psi_{\pm}\rangle=\dfrac{1}{\sqrt{2}}(i^n|n\rangle_a|0\rangle_b\pm
|0\rangle_a|n\rangle_b)\;,
\end{equation}
which is the entangled number state.

We next consider the case where the ion is in the ground state
$|\downarrow\rangle$ and the two c.m. modes are prepared in
coherent states $|\alpha\rangle_a$ and $|\beta\rangle_b$,
respectively,
\begin{equation}
|\Psi(t=0)\rangle=|\downarrow\rangle|\alpha\rangle_a|\beta\rangle_b
\;. \end{equation} If the interactions (1) and (2) are applied in
succession and the interaction times are chosen again as
$\Omega_1t_1=\Omega_2t_2=\pi/4$, the system evolves to
\begin{eqnarray}
|\Psi(t=t_1+t_2)\rangle
=&\dfrac{|\uparrow\rangle+|\downarrow\rangle}{2}
|i\beta\rangle_a|i\alpha\rangle_b \nonumber\\
&+\dfrac{|\downarrow\rangle-|\uparrow\rangle}{2}
|\alpha\rangle_a|\beta\rangle_b \;.
\end{eqnarray}
Upon detection of the ion state, the state vector collapses to $$
|\Psi_{\pm}\rangle=|\alpha\rangle_a|\beta\rangle_b\pm|i\beta\rangle_a|i\alpha\rangle_b
\eqno{(10)} $$ which is the entangled coherent state or the
two-mode Schr\"{o}dinger cat state. If the interaction times $t_1$
and $t_2$ are chosen as $\Omega_1t_1=\Omega_2t_2=\dfrac{\pi}{2}$,
the initial state (8) evolves to
\begin{eqnarray}
|\Psi(t=t_1 +t_2)
\rangle=&\dfrac{|\uparrow\rangle+|\downarrow\rangle}{2}|-\alpha\rangle_a|-\beta\rangle_b
\nonumber\\
&+\dfrac{|\downarrow\rangle-|\uparrow\rangle}{2}|\alpha\rangle_a|\beta\rangle_b\;.
\end{eqnarray}
If the atom is detected in the ground state or the excited state,
the state vector collapses to
\begin{equation}
|\Psi_{\pm}\rangle=|-\alpha\rangle_a|-\beta\rangle_b\pm|\alpha\rangle_a|\beta\rangle_b
\;,
\end{equation}
which is also the entangled coherent state.

Finally we consider the case when the ion is prepared in the
symmetric state
$|+\rangle=(|\uparrow\rangle+|\downarrow\rangle)/\sqrt{2}$ or the
antisymmetric state
$|-\rangle=(|\uparrow\rangle-|\downarrow\rangle)/\sqrt{2}$, and
the two c.m. modes are prepared in either $|0\rangle$ or
$|1\rangle$. When the two interactions (1) and (2) are applied in
succession with $\Omega_1t_1=\Omega_2t_2=\pi/4$, we obtain
\begin{eqnarray}
|-\rangle|i\rangle_a|j\rangle_b\longrightarrow|-\rangle|i\rangle_a|j\rangle_b
~~~~~(i,j=0,1) \;,\nonumber\\
|+\rangle|0\rangle_a|0\rangle_b\longrightarrow|+\rangle|0\rangle_a|0\rangle_b\;,
\nonumber\\
|+\rangle|0\rangle_a|1\rangle_b\longrightarrow{i}|+\rangle|1\rangle_a|0\rangle_b\;,
\nonumber\\
|+\rangle|1\rangle_a|0\rangle_b\longrightarrow{i}|+\rangle|0\rangle_a|1\rangle_b\;,
\nonumber\\
|+\rangle|1\rangle_a|1\rangle_b\longrightarrow{-1}|+\rangle|1\rangle_a|1\rangle_b\;,
\end{eqnarray}
where the state to the left of the arrow represents the initial
state and the state to the right is the state at $t=t_1 +t_2 $.
Eq. (13) represents a Fredkin gate operation\cite{Fredkin}, apart
from phase factors that can be eliminated by an appropriate
setting of the phase of subsequent logic
operations\cite{Wineland}.

In summary, we have presented an efficient scheme that  generates
SU(2) Schr\"odinger cat states, entangled number states, and
entangled coherent states and that gives a Fredkin gate operation
in a trapped ion. It should be pointed out that our proposed
scheme for generating nonclassical motional states is not
deterministic, because the generation is accomplished via post
selection, i.e., selective measurement of a product state after
the state has been produced. Whether the final motional state we
obtain is $|\Psi_{+}\rangle$ or $|\Psi_{-}\rangle$ of Eq. (5) [the
argument applies equally to Eqs. (7), (10), or (12)] depends upon
the outcome of the measurement of the internal state of the ion.
When a particular state $|\Psi_{+}\rangle$ (or $|\Psi_{-}\rangle$)
is desired, our method cannot always succeed; the probability of
success is 50\%. It should also be pointed out that our analysis
assumes a perfectly isolated system. In a real experiment, the
quantum motion of a trapped ion is obviously limited by sources of
decoherence. One most important source is the heating of the
motion of the ions, which appears to be due to the ambient
fluctuating electrical field in the trap. There have been several
theoretical investigations on the heating of trapped
ions\cite{Lamoreaux,James,SM}. In a recent experiment in a Paul
trap\cite{Roos}, the heating rate has been measured to be
sufficiently low that, for relatively large traps, heating may not
be a critically limiting process. The ability to engineer quantum
state as well as the small heating rates in experiments\cite{Roos}
lead us to hope that our present scheme may be realized
experimentally in a near future.

This research was supported by the Brain Korea 21 Project of the
Korean Ministry of Education, by the Korea Science and
Engineering Foundation under Contract No. 1999-2-121-005-3, and
by the Korea Atomic Energy Research Institute.

\end{document}